\documentclass[showaps,graphicx,twocolumn]{revtex4}
\usepackage{amssymb}
\usepackage{amsmath}
\usepackage{graphicx}
\usepackage{array}
\usepackage{multirow}
\usepackage{subfigure}
\usepackage[colorlinks,
linkcolor=blue,
anchorcolor=blue,
citecolor=blue,
urlcolor=blue
]{hyperref}
\begin{document}

\title{Refined quantum gates for $ \Lambda $-type atom-photon hybrid systems}

\author{Fang-Fang Du}
\altaffiliation{Duff@nuc.edu.cn}
\author{Yi-Ming Wu}
\author{Gang Fan}

\affiliation{
   Science and Technology on Electronic Test and Measurement Laboratory, North University of China, Taiyuan 030051, China
}

\begin{abstract}
High-efficiency quantum information processing is equivalent to the fewest quantum resources and the simplest operations by means of logic qubit gates. Based on the reflection geometry of a single photon interacting with a three-level $\Lambda$-typle atom-cavity system,
we present some refined protocols
for realizing controlled-not (CNOT), Fredkin, and Toffoli gates on hybrid systems. The  first control qubit of our gates is
encoded on a flying photon, and the rest qubits are encoded on the atoms in optical cavity. Moreover,
these quantum gates can be extended to the optimal synthesis of multi-qubit CNOT, Fredkin and Toffoli gates with $ O(n) $ optical elements without auxiliary photons or atoms. Further, the simplest single-qubit operations are applied to the photon only,
which make these logic gates experimentally feasible with current technology.
\end{abstract}

\maketitle

\section{Introduction}      \label{sec1}
The evolution of quantum computing and quantum communication are inseparable from the construction and improvement of basic framework, where the logical qubit gate is one of the primary modules \cite{Elementary1995}.
By far, plentiful and various logical gates are presented with diverse systems, such as linear optics \cite{Nearly2004,Linear-optics2006, A2016, A2019}, quantum dots (QDs) \cite{Deterministic2013,Wang_2014}, cavity-atom
systems \cite{Deterministic_2010,atom2014,Hyperparallel2016 }, cross-Kerr nonlinearity \cite{Constructing2018}, ion trap \cite{Microwave2011} and nitrogen-vacancy center \cite{Universal2015}.
Photon possesses available single-qubit operations, easy and intuitional measure, low decoherence, faithful and efficient transmission of information.
In addition, the photon is labeled as the best flying qubits for quantum information processing (QIP), i.e., quantum teleportation \cite{Teleporting1993}, quantum secure direct communication \cite{Li_2019, Li_2020, Practical2021, Drastic2021, One-step2022}, and quantum repeaters \cite{Yan_2021} to connect the two adjacent nodes in quantum networks.
Nemoto \emph{et al}. \cite{Nearly2004} proposed a controlled-NOT (CNOT) gates with the aid of one ancillary photon.
Fiur\'{a}\v{s}ek \emph{et al}. \cite{Linear-optics2006} presented the linear three-qubit Toffoli (controlled-controlled-NOT) and Fredkin (controlled-swap) gates with auxiliary photons.
Patel \emph{et al.} \cite{A2016} experimentally implemented a Fredkin gate by adding a control qubit to the swap unitary.
Besides, multi-qubit gates are favored in scalable QIP because of fertilizing applicability and reducing resource cost.

Using fewer resources and more simplified quantum circuits has always been the priority of logic qubit gates.
In 2007, Ralph \emph{et al.} \cite{Efficient2007} took advantage of the higher-dimensional Hilbert spaces to form qudit for constructing a theoretical Toffoli gate.
Based on this multi-level physical systems, Liu and Wei \cite{Liu_2020} reduced the complexity of Fredkin gate assisted by five CNOT gates.
In 2010, Bonato  \emph{et al}. \cite{CNOT2010} designed  the quantum logical CNOT gates. In 2013, Wei and Deng \cite{WEIDENG2013} proposed hybrid universal quantum gates assisted by stationary electron spins of a QD in a double-sided microcavity.
In order to improve the ability of information processing, hyper-parallel CNOT, Toffoli gates with the QED have been presented\cite{Deterministic2013,Wei2016}.
Li and Long \cite{Hyperparallel2016} generalized hyper-CNOT$ ^{n} $ gate based on the effective
Kerr nonlinear in double-sided optical cavities.
Ref. \cite{LIDENG2016} proposed a recycling quantum entangling gate to perfect the errors with QD in low-$Q$ optical microcavity.

The weak interaction between photons hinders the photonic control of the quantum gates.
A single atom or artificial atom trapped in an optical cavity has become an important platform for the realization of QIP.
The system based on cavity quantum electrodynamics (QED) can be used to exchange information
between static and flying qubits \cite{Repeaters1998}, and a measurable nonlinear phase shift between single photons can be achieved even in the bad-cavity regime.
People have focus on the cavity-atom hybrid gates assisted by $ \Lambda  $-type system, in which the transition between the ground states $ |g_{v}\rangle $ $(  |g_{h}\rangle )$ and the excited state $ |e\rangle $ coupled resonantly to one single-sided optical cavity has been confirmed. Koshino \emph{et al}. \cite{Deterministic_2010} presented a deterministic photon-photon $ \sqrt{\mathrm{SWAP}} $ gate  as temporary memory for a photonic qubit.
Besides, Song \emph{et al}. \cite{Song_2009} proposed two deterministic schemes to implement quantum swap gate and Fredkin
gate encoded for atomic qubits through the input-output process of the cavity.

In this paper, we construct the quantum circuits of three basic logic qubit gates, including two-qubit CNOT gate, three-qubit Fredkin and Toffoli gates, assisted by the a $ \Lambda $-type three-level atom embedded in optical cavity.
The (first) control qubit are all photons that firstly converted into two spatial modes to complete the subsequent circuits by using the concept of qudit. All the qubit operations are employed on the photon only,
and the three hybrid gates can be synthesized without any auxiliary photons, in terms of simplicity.
Besides, the quantum circuits can be extended flexibly to realize the multi-qubit gates.

\section{ Optimal syntheses of  quantum logic gates}    \label{sec2}

\begin{figure}[htp]
	\centering
	\includegraphics[width=0.7\linewidth]{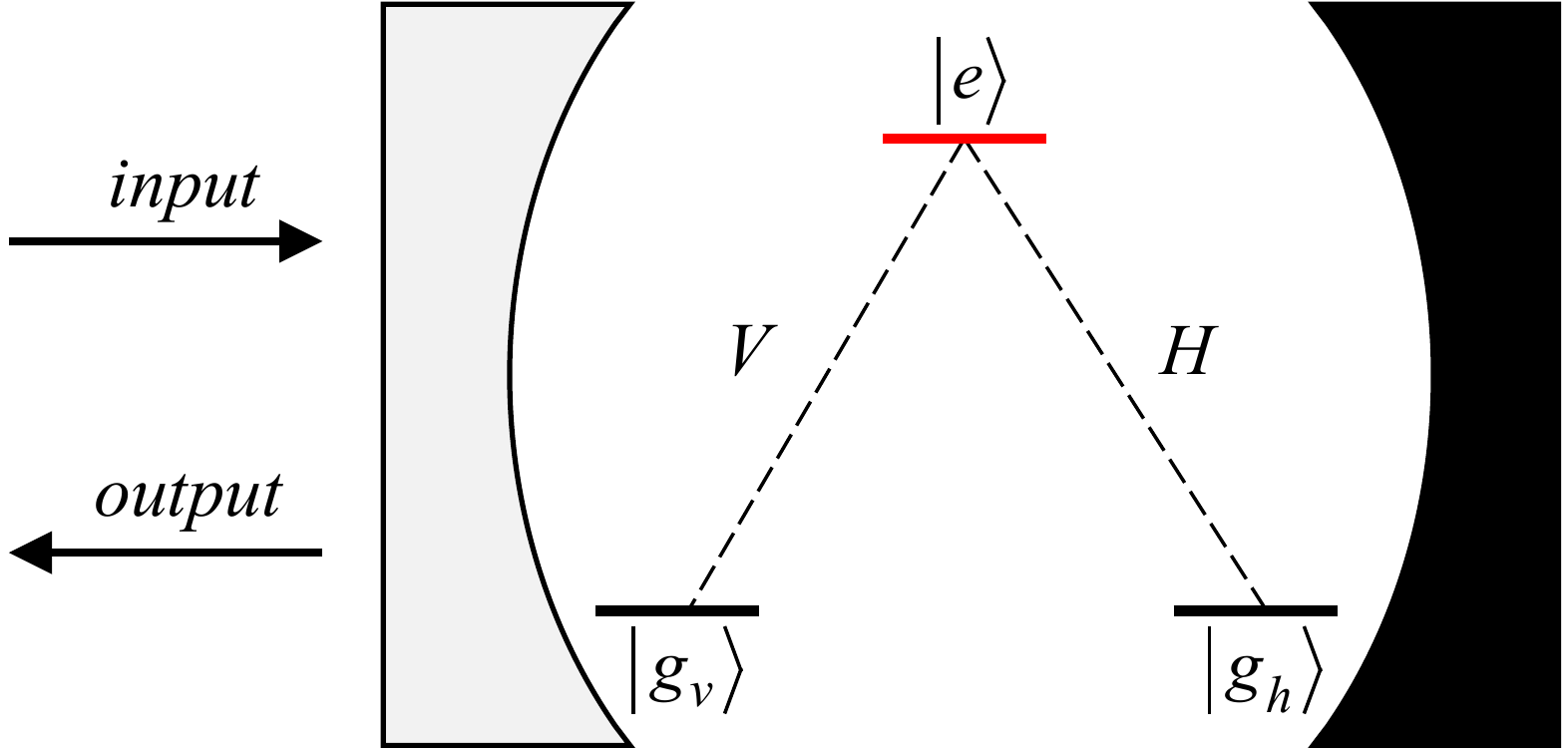}
	\caption{Schematic diagram of the interaction between a photon and a $ \Lambda $-type atom embedded in single-sided optical cavity.
	}\label{fig1}
\end{figure}
An atom considered as a $ \Lambda $-type three-level system shown in Fig. \ref{fig1} is trapped into a single-sided optical cavity.
The ground states ($ |g_{h}\rangle $ and $ |g_{v}\rangle $) and the excited state ($ |e\rangle $) are used to complete the transition $ |g_{h}\rangle \rightarrow |e\rangle $ or $ |g_{v}\rangle \rightarrow |e\rangle $ coupled resonantly to the cavity mode $c_{h}$ or $c_{v}$, which only can be excited by the polarized state $ |H\rangle $ or $ |V\rangle $ of a single photon.
Under the rotating approximation, the Hamiltonian of the whole system 
(limiting $ \hbar=1 $) \cite{Quantum-information2009}
\begin{eqnarray}\label{eq1}
	\hat{H}\!\!\!&=&\!\!\!i\sum_{k=h,v}[\eta_{k}(c_{k}|g_{k}\rangle\langle e|-c_{k}^{\dag}|e\rangle\langle g_{k}|)]-i\frac{\gamma}{2}|e\rangle\langle e| \nonumber\\ &&\!\!\!+i\sqrt{\kappa/2\pi}\int d\omega[b_{k}^{\dag}(\omega)c_{k}-b_{k}(\omega)c_{k}^{\dag}]\nonumber\\
&&\!\!\!+\sum_{k=h,v}\int d\omega b_{k}^{\dag}(\omega)b_{k}(\omega).
\end{eqnarray}
$ c^{\dag}_{k} $ and $ c_{k} (k=h,v)$ represent the creation and annihilation operators of the cavity mode, respectively. The parameters $ b_{k}(\omega) $ and $ b_{k}^{\dag}(\omega) $ are the annihilation and creation operators of the corresponding polarized states, respectively. $ \kappa $ is the damping rate of the cavity system. $ \gamma $ is the spontaneous rate of the excited state $ |e\rangle $. The parameter $ \eta_{k} $ is the coupling strength between the atom transition and the cavity mode.

Suppose that the atom is in the ground state $ |g_{k}\rangle $ and the input photon is in the corresponding $K$-polarized state ($ K = H, V $), the general time-based wave function of the system can be described as \cite{Modification1995,Probing2006,Complete20161}
\begin{eqnarray}\label{eq3}
	|\psi(t)\rangle&=&\sum_{k=h,v}\bigg( \int_{-\infty}^{+\infty}d\omega\beta_{k,\omega}(t)b_{k}^{\dag}|g_{k},0,vac\rangle  \nonumber\\
	&&+\beta_{e}(t)|e,0,vac\rangle+\beta_{g,k}(t)|g_{k},1,vac\rangle\bigg).
\end{eqnarray}
Here, $ |vac\rangle $ represents the vacuum state in the fiber mode, 0 or 1 denotes the number of the photon in the $K$-polarized state. According to the Heisenberg equations of motion and the standard input-output relation, we can obtain the parameters $ \beta_{k,\omega}^{out} $, $ \beta_{g,k} $ and $ \beta_{e} $. The relation can be illustrated by \cite{Deterministic_2010,Song_2009}
\begin{eqnarray}\label{eq4}
	\dot{\beta}_{g,k}(t)&=&-\eta_{k}\beta_{e}-\sqrt{\kappa}\beta_{k,\omega}^{in}-\frac{\kappa}{2}\beta_{g,k}(t),\nonumber\\
	\dot{\beta}_{e}(t)&=&\left[ \eta_{v}\beta_{g,v}(t)+\eta_{h}\beta_{g,h}(t)\right] -\frac{\gamma}{2}\beta_{e}(t),\nonumber\\
	\beta_{k,\omega}^{out}&=&\beta_{k,\omega}^{in}+\sqrt{\kappa}\beta_{g,k}(t),
\end{eqnarray}
where $ \beta_{k,\omega}^{in}=\frac{1}{\sqrt{2\pi}}\int_{-\infty}^{+\infty}e^{-i\omega t}\beta_{k,\omega}(0)$ is is the amplitude of the input pulse and $\beta_{k,\omega}^{out}$ is the amplitude of the output pulse. By linearizing the Eq. (\ref{eq1}),  $\beta_{k,\omega}^{out}$ can be solved by Fourier transform, shown in Eq. (\ref{eq5}),
\begin{eqnarray}\label{eq5}
	\beta_{k,\omega}^{out}\!\!\!\!\!\!&=&\!\!\!\!\!\!\beta_{k,\omega}^{in}-\kappa\frac{\left[ \eta_{\bar{k}}^{2}+\left(i\omega+\frac{\kappa}{2}\right)\left(i\omega+\frac{\gamma}{2}\right)\right]\beta_{k,\omega}^{in}-\eta_{k}\eta_{\bar{k}}\beta_{\bar{k},\omega}^{in}}{\left(i\omega+\frac{\kappa}{2}\right)\left[\eta_{k}^{2}+\eta_{\bar{k}}^{2}+\left(i\omega+\frac{\kappa}{2}\right)\left(i\omega+\frac{\gamma}{2}\right) \right] }\nonumber\\
	\!\!\!\!\!\!&=&\!\!\!\!\!\!r_{h1}\beta_{k,\omega}^{in}+r_{h2}\beta_{\bar{k},\omega}^{in},\nonumber\\
	r_{h1}\!\!\!\!\!\!&=&\!\!\!\!\!\!\frac{1}{\left(i\omega+\frac{\kappa}{2}\right)}\left[ \left(i\omega-\frac{\kappa}{2}\right)\right.\nonumber\\
	&&\!\!\!\!\!\!\left.+\frac{\kappa\eta_{k}^{2}}{\eta_{k}^{2}+\eta_{\bar{k}}^{2}+\left(i\omega+\frac{\kappa}{2}\right)\left(i\omega+\frac{\gamma}{2}\right) }\right],\nonumber\\
	r_{h2}\!\!\!\!\!\!&=&\!\!\!\!\!\!\frac{\kappa\eta_{k}\eta_{\bar{k}}}{\left(i\omega+\frac{\kappa}{2}\right)\left[\eta_{k}^{2}+\eta_{\bar{k}}^{2}+\left(i\omega+\frac{\kappa}{2}\right)\left(i\omega+\frac{\gamma}{2}\right) \right]},
\end{eqnarray}
where $ \{k,\bar{k}\}\in\{h,v\} $ and $ \bar{k}\neq k$.
When the atom does not couple to the input filed for an cold cavity, i.e., the coupling strength $ \eta_{h}=\eta_{v}=0 $, the amplitude of the output pulse $\beta_{k,\omega,0}^{out}$ is
\begin{eqnarray}\label{eq6} 
	\beta_{k,\omega,0}^{out}=r_{0}\beta_{k,\omega}^{in},\quad r_{0}=\frac{i2\omega-\kappa}{i2\omega+\kappa}.
\end{eqnarray}

In the case $ \kappa\gg\gamma$, $ \omega $, we can obtain
$ \beta_{k,\omega,0}^{out}=-\beta_{k,\omega}^{in} $ ($r_{0}\rightarrow -1$) with a $\pi$ phase. When $ \eta_{k}^{2}\gg\kappa\gamma $, we can obtain $ \beta_{k,\omega}^{out}=\beta_{\bar{k},\omega}^{in} $ ($r_{h1}\rightarrow 0, r_{h2}\rightarrow1$) with the bit flips of the states of the atom and the photon in Eq. (\ref{eq5}). In detail, if the atom is in the state $ |g_{h}\rangle (|g_{v}\rangle) $, the resonance between the cavity-atom system and the photon in the polarized state $ |H\rangle (|V\rangle) $ fells a hot cavity, both the states of the photon and  atom will perform qubit-flip operations. Otherwise, if the atom is in the state $ |g_{v}\rangle (|g_{h}\rangle) $, the photon in the polarized state $ |H\rangle (|V\rangle) $ will encounter a cold cavity, and thus the photon will be reflected with a phase-flip operation. 
The above detailed rules of interaction can be summarized as follows
\begin{eqnarray}\label{eq7}
	|H\rangle|g_{h}\rangle&\rightarrow&|V\rangle|g_{v}\rangle,\nonumber\\
	|H\rangle|g_{v}\rangle&\rightarrow&-|H\rangle|g_{v}\rangle,\nonumber\\
	|V\rangle|g_{h}\rangle&\rightarrow&-|V\rangle|g_{h}\rangle,\nonumber\\
	|V\rangle|g_{v}\rangle&\rightarrow&|H\rangle|g_{h}\rangle.
\end{eqnarray}

\begin{figure}[htp]
	\centering
	\includegraphics[width=0.8\linewidth]{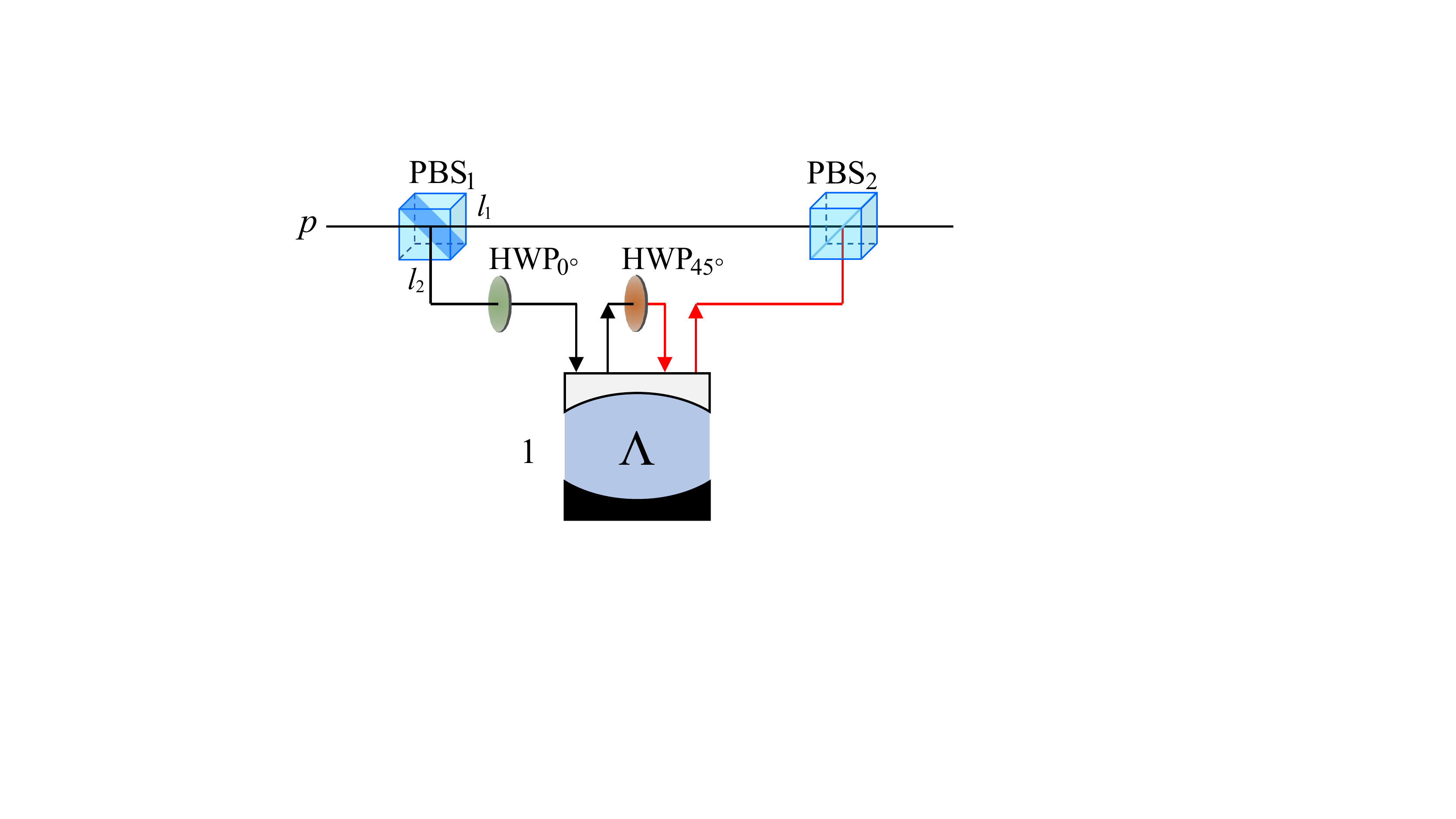}
	\caption{Schematic diagram of a two-qubit hybrid CNOT gate.
		PBS$ _{m} $ (m=1,2) is polarized beam splitter which transforms the horizontal polarized state $ |H\rangle $ of a photon and reflects the vertical polarized state $ |V\rangle $ of a photon.
		HWP$ _{j^{\circ}} $ is a half-wave plate (HWP) oriented at $ j$ degrees, and HWP$ _{0^{\circ}} $ and HWP$ _{45^{\circ}} $ represent a phase-flip operation ($ \sigma _{z}$) and a qubit-flip operation ($ \sigma _{x} $), respectively, where $ \sigma _{z}=|H\rangle\langle H|-|V\rangle\langle V|$ and $ \sigma_{x}=|H\rangle\langle V|+|V\rangle\langle H|$.
	}\label{fig2}
\end{figure}

By taking advantage of the $ \Lambda $-type cavity-atom model, we can construct basic qubit gates, including two-qubit CNOT gate, three-qubit Fredkin gate and Toffoli gate.
The CNOT gate consists of one control qubit and one target qubit, where the photon $ p $ is the control qubit and the atom $ 1 $ is the target qubit.
The concrete quantum circuit of the hybrid CNOT gate is shown in Fig. \ref{fig2}. Suppose that the initial states of one photon $ p $ and one atom $ 1 $ are $ (\alpha_{1}|H\rangle+\alpha_{2}|V\rangle)_{p}$ and $(\zeta_{1}|g_{h}\rangle+\zeta_{2}|g_{v}\rangle)_{1} $, respectively. The complex coefficients comply with the normalization theorem $ |\alpha_{1}|^{2}+|\alpha_{2}|^{2}=1 $ and $|\zeta_{1}|^{2}+|\zeta_{2}|^{2}=1  $.

	First, the photon $ p $ passes through the first polarized beam splitter (PBS$ _{1} $), which transmits polarized state $ |H\rangle $ of the photon into line $l_{1}$ and reflects polarized state $ |V\rangle $ of the photon into line $l_{2}$. The photon on line $l_{2}$ passes through HWP$ _{0^{\circ}} $, which denotes a half-wave plate (HWP) oriented at $ 0^\circ$ degrees to complete a $ \sigma _{z}$ operation, in which  $ \sigma _{z}=|H\rangle\langle H|-|V\rangle\langle V|$, and then interacts with the atom in the single-sided cavity using the rules described in the Eq. (\ref{eq7}). The whole state of the hybrid system evolves into
	\begin{eqnarray}\label{eq8}
		|\phi\rangle_{1}\!\!\!\!\!\!&=&\!\!\!\!\!\!
		\alpha_{1}\zeta_{1}|H\rangle_{l_{1}}|g_{h}\rangle_{1}
		+\alpha_{1}\zeta_{2}|H\rangle_{l_{1}}|g_{v}\rangle_{1}\nonumber\\
		&&\!\!\!\!\!\!+\alpha_{2}\zeta_{1}|V\rangle_{l_{2}}|g_{h}\rangle_{1}
		-\alpha_{2}\zeta_{2}|H\rangle_{l_{2}}|g_{h}\rangle_{1}.
	\end{eqnarray}
	
	Second, the photon on line $l_{2}$ passes through  HWP$ _{45^\circ} $, which represents a qubit-flip operation ($ \sigma _{x} $), i.e., $ \sigma_{x}=|H\rangle\langle V|+|V\rangle\langle H|$. The whole state of this system is changed into
	\begin{eqnarray}\label{eq9}
		|\phi\rangle_{2}\!\!\!\!\!\!&=&\!\!\!\!\!\!
		\alpha_{1}\zeta_{1}|H\rangle_{l_{1}}|g_{h}\rangle_{1}
		+\alpha_{1}\zeta_{2}|H\rangle_{l_{1}}|g_{v}\rangle_{1}\nonumber\\
		&&\!\!\!\!\!\!+\alpha_{2}\zeta_{1}|H\rangle_{l_{2}}|g_{h}\rangle_{1}
		-\alpha_{2}\zeta_{2}|V\rangle_{l_{2}}|g_{h}\rangle_{1}.
	\end{eqnarray}
	
	Finally,   the photon on line 2 interacts with the atom again, and  both lines $l_{1}$ and $l_{2}$ converge on another PBS$ _{2} $, thereby evolving to
	\begin{eqnarray}\label{eq10}
		|\phi\rangle_{3}\!\!\!\!\!\!&=&\!\!\!\!\!\!
		\alpha_{1}|H\rangle_{p}(\zeta_{1}|g_{h}\rangle
		+\zeta_{2}|g_{v}\rangle)_{1}\nonumber\\
		&&\!\!\!\!\!\!+\alpha_{2}|V\rangle_{p}(\zeta_{2}|g_{h}\rangle
		+\zeta_{1}|g_{v}\rangle_{1}.	\end{eqnarray}
	
	Based on Eqs. (\ref{eq8})-(\ref{eq10}), one can find that the hybrid CNOT gate is refined, low cost, and maneuverable, as the system can be constructed without any auxiliary particle, and meanwhile the single-qubit operations are only applied  to the photon.
	
	\begin{figure}[htp]
		\centering
		\includegraphics[width=0.7\linewidth]{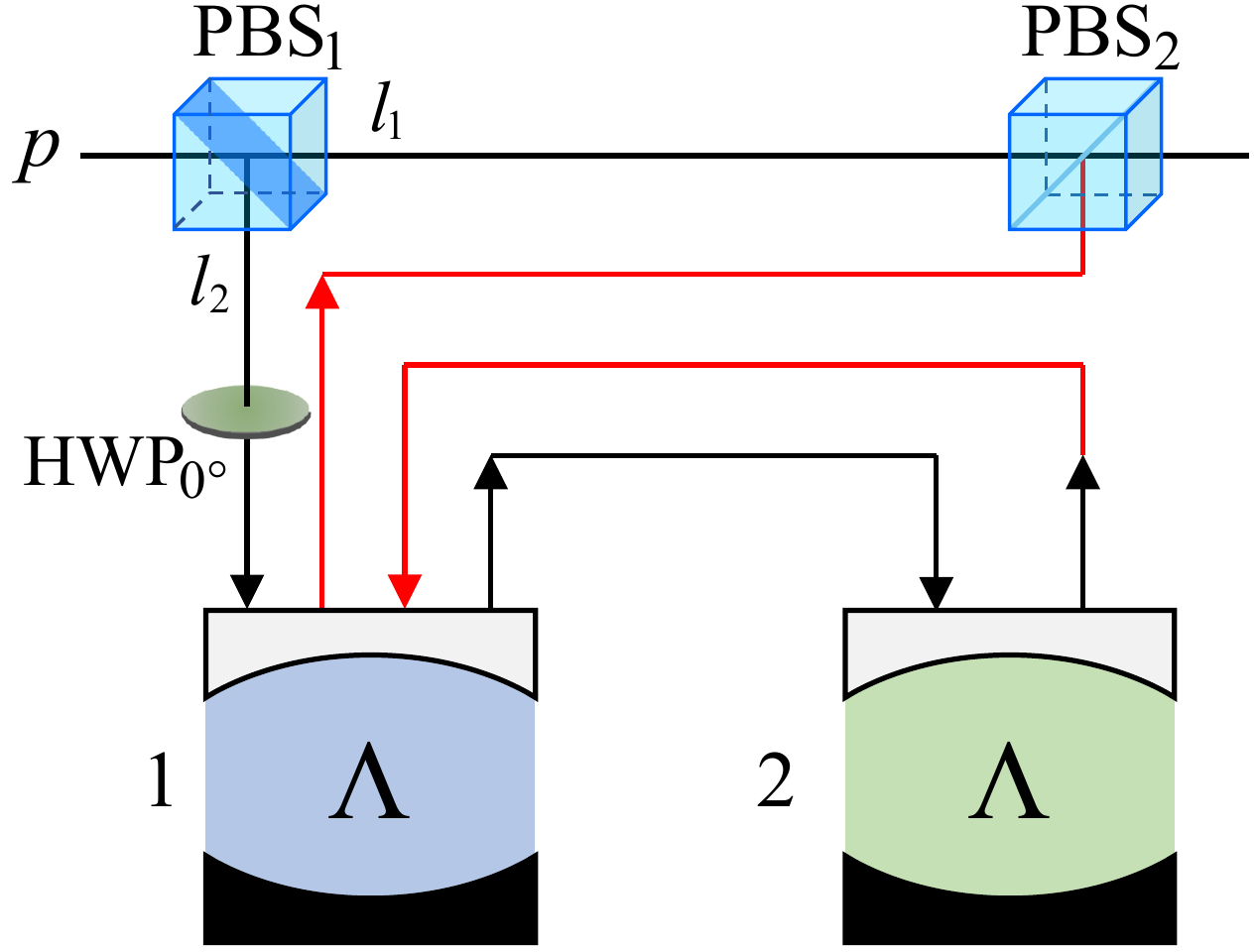}
		\caption{Schematic diagram of a three-qubit hybrid Fredkin gate.}\label{fig3}
	\end{figure}
	In addition, a hybrid three-qubit Fredkin gate can be synthesized as shown in Fig. \ref{fig3}, where the photon $ p $ is the control qubit and two atoms $ 1 $ and $ 2 $ are the target qubits.
Suppose that the initial state of the hybrid system also conforms to the normalization theorem, and the specific state is
	\begin{eqnarray}\label{eq11}
		|\Phi\rangle_{0}\!\!\!\!\!\!&=&\!\!\!\!\!\!(\alpha_{1}|H\rangle+\alpha_{2}|V\rangle)_{p}\otimes (\zeta_{1}|g_{h}\rangle+\zeta_{2}|g_{v}\rangle)_{1}\nonumber\\
		&&\!\!\!\!\!\!\otimes(\varepsilon_{1}|g_{h}\rangle+\varepsilon_{2}|g_{v}\rangle)_{2}.
	\end{eqnarray}

	First, the photon $ A $ passes through PBS$ _{1}  $, and then the polarized state $|H\rangle$ of the photon is directly transmitted into line $l_{1}$, while the other polarized state $|V\rangle$ of the photon is reflected into line $l_{2}$ to pass through HWP$ _{0^{\circ}} $ and interacts with the atom $ 1 $ in the cavity (black lines), the whole state of the system changes to
	\begin{eqnarray}\label{eq12}
		|\Phi\rangle_{1}\!\!\!\!\!\!&=&\!\!\!\!\!\![\alpha_{1}|H\rangle_{l_{1}}\otimes (\zeta_{1}|g_{h}\rangle+\zeta_{2}|g_{v}\rangle)_{1}+\alpha_{2}|V\rangle_{l_{2}}\zeta_{1}|g_{h}\rangle_{1}\nonumber\\
		&&\!\!\!\!\!\!-\alpha_{2}|H\rangle_{l_{2}}\zeta_{2}|g_{h}\rangle_{1}]\otimes(\varepsilon_{1}|g_{h}\rangle+\varepsilon_{2}|g_{v}\rangle)_{1}.
	\end{eqnarray}
	
	Second, the photon on line $l_{2}$ interacts with the atom $ 2 $ in another cavity (black lines), thereby becoming to
	\begin{eqnarray}\label{eq13}
		|\Phi\rangle_{2}\!\!\!\!\!\!&=&\!\!\!\!\!\!\alpha_{1}|H\rangle_{l_{1}}\otimes (\zeta_{1}|g_{h}\rangle+\zeta_{2}|g_{v}\rangle)_{1}(\varepsilon_{1}|g_{h}\rangle+\varepsilon_{2}|g_{v}\rangle)_{2}\nonumber\\
		&&\!\!\!\!\!\!-\alpha_{2}|V\rangle_{l_{2}}[\zeta_{1}|g_{h}\rangle\varepsilon_{1}|g_{h}\rangle
		+\zeta_{2}|g_{h}\rangle\varepsilon_{1}|g_{v}\rangle]_{12}	\nonumber\\
		&&\!\!\!\!\!\!+\alpha_{2}|H\rangle_{l_{2}}[\zeta_{1}|g_{h}\rangle\varepsilon_{2}|g_{h}\rangle
		+\zeta_{2}|g_{h}\rangle\varepsilon_{2}|g_{v}\rangle]_{12}.
	\end{eqnarray}
	
	Finally, the photon on line  $l_{2}$  interacts with the atom $ 1 $  again (red lines). After that, both the two lines converge on the another PBS$ _{2} $ and therefore the whole state of this system is induced into
	\begin{eqnarray}\label{eq14}
		|\Phi\rangle_{3}\!\!\!\!\!\!&=&\!\!\!\!\!\!\alpha_{1}|H\rangle_{p} (\zeta_{1}|g_{h}\rangle+\zeta_{2}|g_{v}\rangle)_{1}(\varepsilon_{1}|g_{h}\rangle+\varepsilon_{2}|g_{v}\rangle)_{2}\nonumber\\
		&&\!\!\!\!\!\!+\alpha_{2}|V\rangle_{p} (\varepsilon_{1}|g_{h}\rangle+\varepsilon_{2}|g_{v}\rangle)_{1}(\zeta_{1}|g_{h}\rangle+\zeta_{2}|g_{v}\rangle)_{2}.
	\end{eqnarray}
	Based on Eqs. (\ref{eq11})-(\ref{eq14}), one can see that this system completes the function of the hybrid Fredkin gate, that is, if and only if the polarized state of photon $ p $ is $ |V\rangle $, the two atoms $ 1 $ and 2 swap the qubit information with each other. Our hybrid Fredkin gate requires only two optical elements PBSs and one single-qubit operation on the photon. Likewise, low consumption and easy handling indicate the feasibility of this scheme.

\begin{figure}[htp]
		\centering
		\includegraphics[width=0.8\linewidth]{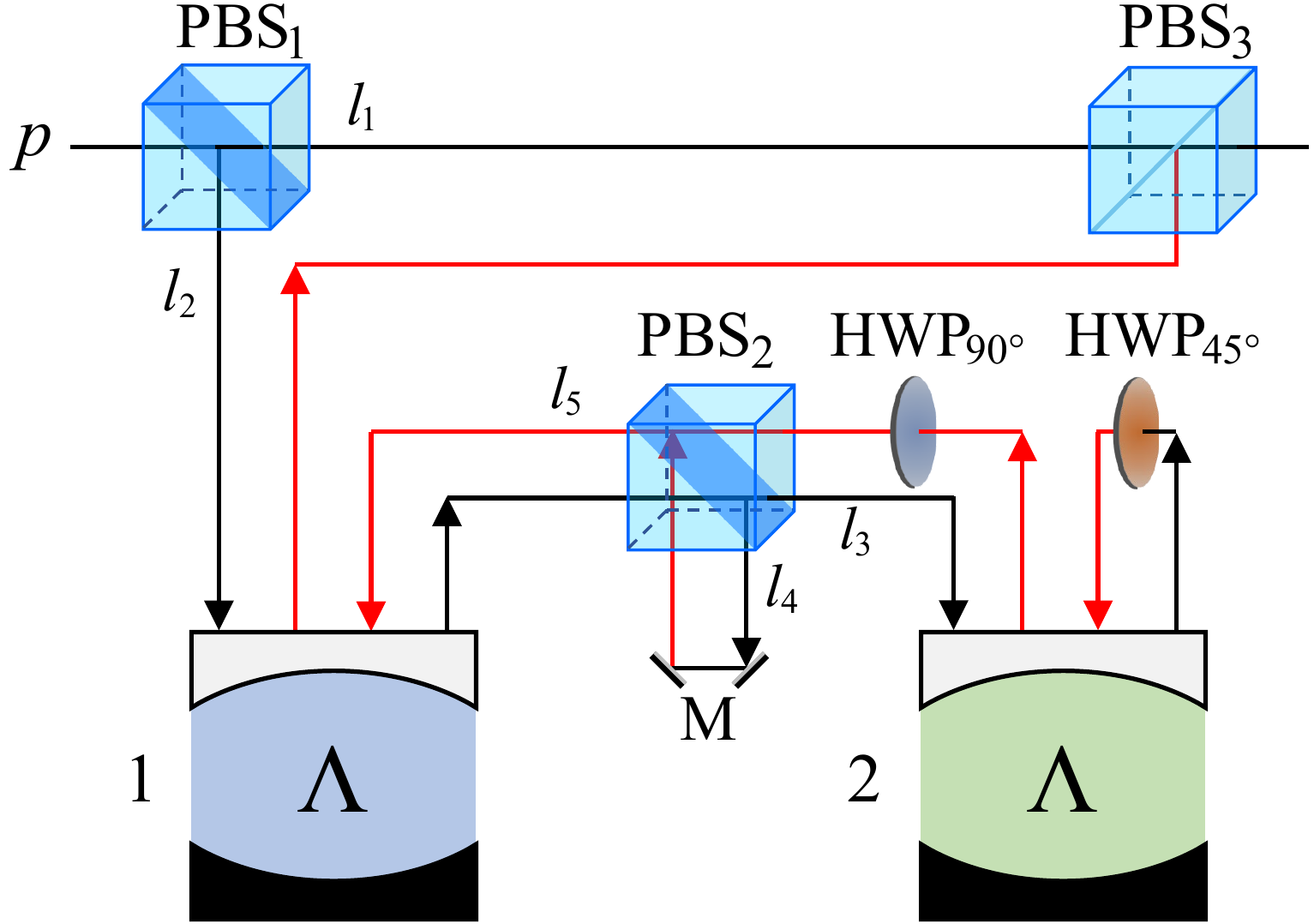}
		\caption{Schematic diagram of a hybrid Toffoli gate. HWP$ _{90^{\circ}} $ implements a $ -\sigma _{x} $ operation. M denotes a mirror.
		}\label{fig4}
	\end{figure}

	 Further, the hybrid Toffoli gate shown in Fig. \ref{fig4} can be formed by taking a photon and two atoms, and suppose that this hybrid system has the same initial state as the hybrid Fredkin gate shown in Eq. (\ref{eq11}).
	Here, we regard the photon $ p $ and the atom $ 1 $ as control qubits and the atom $ 2 $ as target qubit. 
	
	First, the photon $ p $ passes through PBS$ _{1} $, and the polarized state $ |H\rangle $ of the photon is transmitted into line  $l_{1}$ , while the polarized state $ |V\rangle $ of the photon is reflected into line  $l_{2}$  and interacts with the atom $ 1 $ in the first cavity.
Then the photon on line  $l_{2}$  passes through PBS$ _{2} $, likewise, the polarized state $ |V\rangle $ of the photon on line  $l_{2}$  is reflected into line  $l_{4}$, while the polarized state $ |H\rangle $ of the photon is is transmitted into line  $l_{3}$  and interacts with the atom $ 2 $ in the another cavity, thereby the whole state of this hybrid system changes into
	\begin{eqnarray}\label{eq16}
		|\Psi\rangle_{1}\!\!\!\!\!\!&=&\!\!\!\!\!\!\alpha_{1}|H\rangle_{l_{1}} (\zeta_{1}|g_{h}\rangle+\zeta_{2}|g_{v}\rangle)_{1}(\varepsilon_{1}|g_{h}\rangle+\varepsilon_{2}|g_{v}\rangle)_{2}\nonumber\\
		&&\!\!\!\!\!\!-\alpha_{2}|V\rangle_{l_{4}}[\zeta_{1}|g_{h}\rangle\varepsilon_{1}|g_{h}\rangle
		+\zeta_{1}|g_{h}\rangle\varepsilon_{2}|g_{v}\rangle]_{12}	\nonumber\\
		&&\!\!\!\!\!\!-\alpha_{2}|H\rangle_{l_{3}}(\zeta_{2}|g_{h}\rangle\varepsilon_{1}|g_{v}\rangle)_{12}\nonumber\\
		&&\!\!\!\!\!\!
		+\alpha_{2}|V\rangle_{l_{3}}(\zeta_{2}|g_{h}\rangle\varepsilon_{2}|g_{v}\rangle)_{12}.
	\end{eqnarray}
	
	Second, before and after the photon on line  $l_{3}$  interacts with the atom $ 2 $ again, the photon passes through HWP$ _{45^{\circ}} $ and HWP$ _{90^{\circ}} $, respectively, where the latter represents a $ -\sigma_{z} $ operation, i.e., $ -\sigma_{z}=-|H\rangle\langle H|+|V\rangle\langle V| $.
	Meanwhile, the photon on line  $l_{4}$  is reflected by two mirrors, and both lines  $l_{3}$  and  $l_{4}$  converge on PBS$ _{2} $ again (red lines). The state of this hybrid system is induced into	
	\begin{eqnarray}\label{eq17}
		|\Psi\rangle_{2}\!\!\!\!\!\!&=&\!\!\!\!\!\!\alpha_{1}|H\rangle_{l_{1}} (\zeta_{1}|g_{h}\rangle+\zeta_{2}|g_{v}\rangle)_{1}(\varepsilon_{1}|g_{h}\rangle+\varepsilon_{2}|g_{v}\rangle)_{2}\nonumber\\
		&&\!\!\!\!\!\!-\alpha_{2}|V\rangle_{l_{5}}[\zeta_{1}|g_{h}\rangle\varepsilon_{1}|g_{h}\rangle
		+\zeta_{1}|g_{h}\rangle\varepsilon_{2}|g_{v}\rangle]_{12}	\nonumber\\
		&&\!\!\!\!\!\!+\alpha_{2}|H\rangle_{l_{5}}[\zeta_{2}|g_{h}\rangle\varepsilon_{1}|g_{v}\rangle
		+\zeta_{2}|g_{h}\rangle\varepsilon_{2}|g_{h}\rangle]_{12}.
	\end{eqnarray}
	
	Finally, the photon on line 5 interacts with the atom $ 1 $ again. After that, the photon on lines  $l_{1}$  and   $l_{5}$  converge on the PBS$ _{3} $, thereby evolving into
	\begin{eqnarray}\label{eq18}
  |\Psi\rangle_{T}\!\!\!\!\!\!&=&\!\!\!\!\!\!\alpha_{1}|H\rangle_{p} (\zeta_{1}|g_{h}\rangle+\zeta_{2}|g_{v}\rangle)_{m}(\varepsilon_{1}|g_{h}\rangle+\varepsilon_{2}|g_{v}\rangle)_{n}\nonumber\\
  &&\!\!\!\!\!\!+\alpha_{2}|V\rangle_{p}\zeta_{1}|g_{h}\rangle_{m}(\varepsilon_{1}|g_{h}\rangle
  +\varepsilon_{2}|g_{v}\rangle)_{n} \nonumber\\
  &&\!\!\!\!\!\!+\alpha_{2}|V\rangle_{p}\zeta_{2}|g_{v}\rangle_{m}(\varepsilon_{1}|g_{v}\rangle
  +\varepsilon_{2}|g_{h}\rangle)_{n}.
 \end{eqnarray}
	Combined with the above description, if
and only if the two control qubits are in the states $ |V\rangle $ and $ |g_{v}\rangle_{1} $, respectively, the target qubit in the state $(\varepsilon_{1}|g_{h}\rangle
  +\varepsilon_{2}|g_{v}\rangle)_{2}$ performs qubit flip.
Obviously, the hybrid Toffoli gate
has been implemented assisted by three optical elements PBSs and two single-qubit operation on the photon only, which are feasible in experiment for quantum computation and quantum communication.

\section{Discussion and conclusion}

\begin{figure}[htp]
	\centering
	\subfigure[]{
		\begin{minipage}[b]{0.9\linewidth}
			\centering
			\includegraphics[width=1\linewidth]{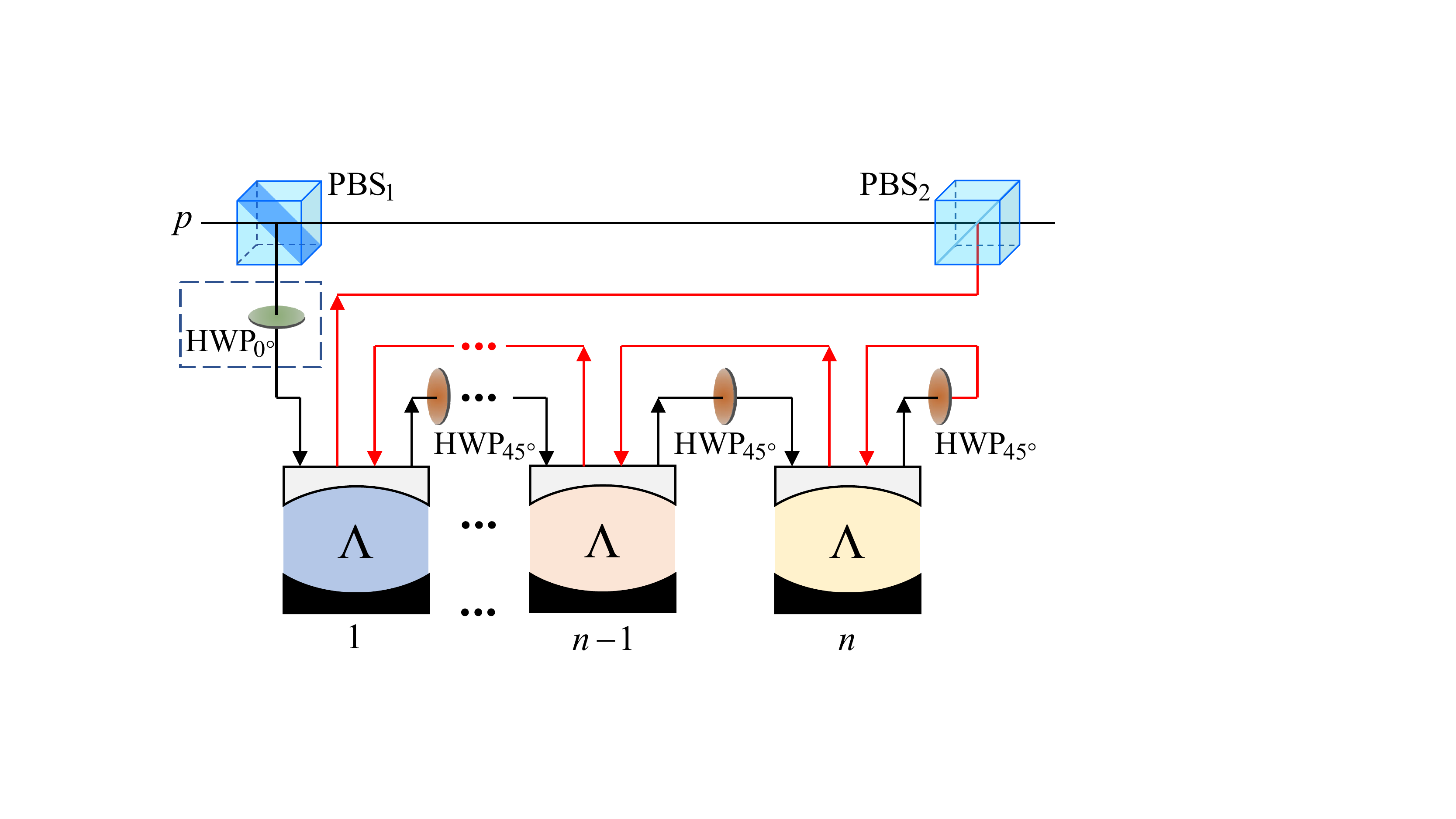}
			\label{fig5a}
	\end{minipage}}	
	\subfigure[]{
		\begin{minipage}[t]{0.9\linewidth}
			\centering
			\includegraphics[width=1\linewidth]{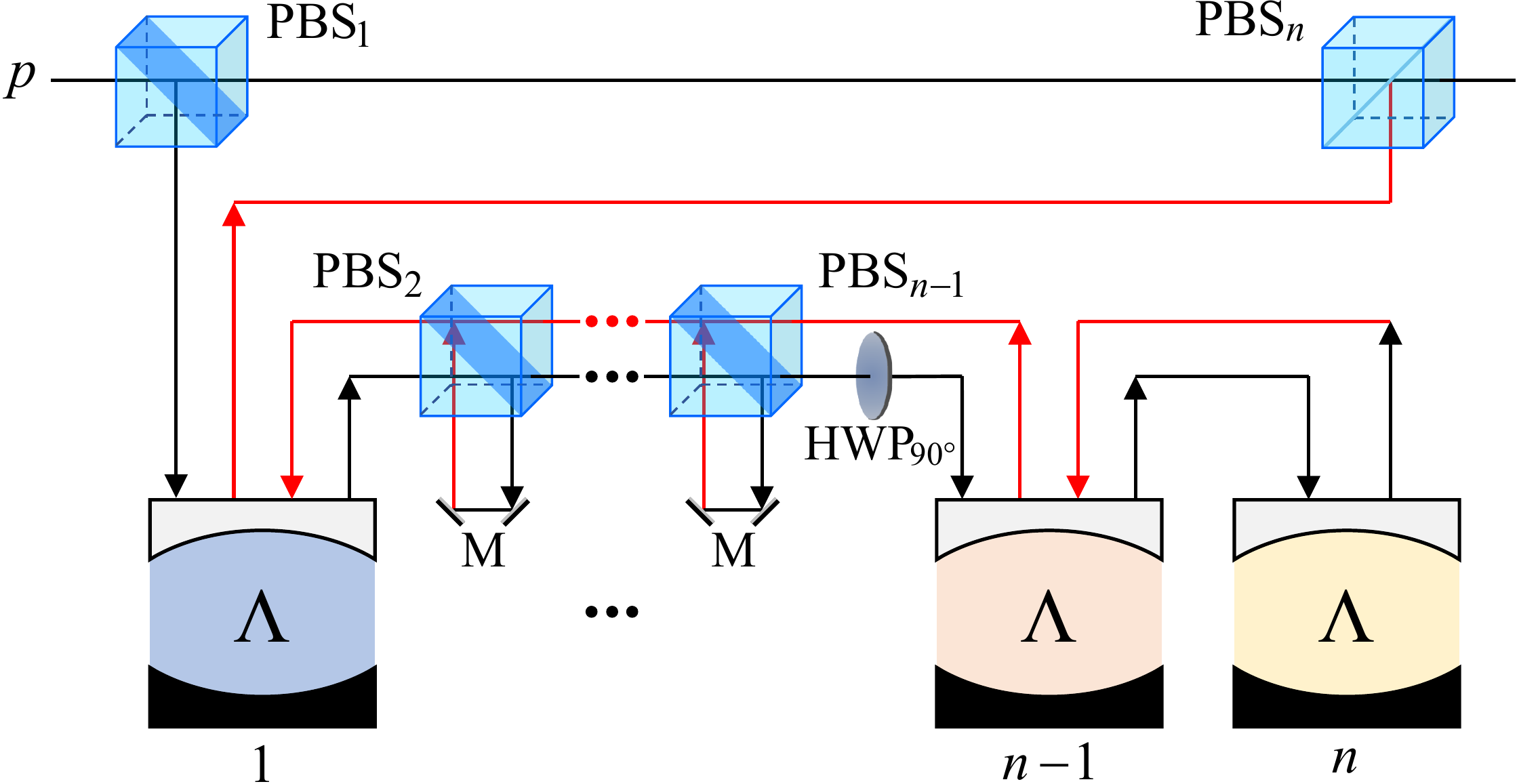}
			\label{fig5b}
	\end{minipage}}
	\subfigure[]{
		\begin{minipage}[t]{0.9\linewidth}
			\centering
			\includegraphics[width=1\linewidth]{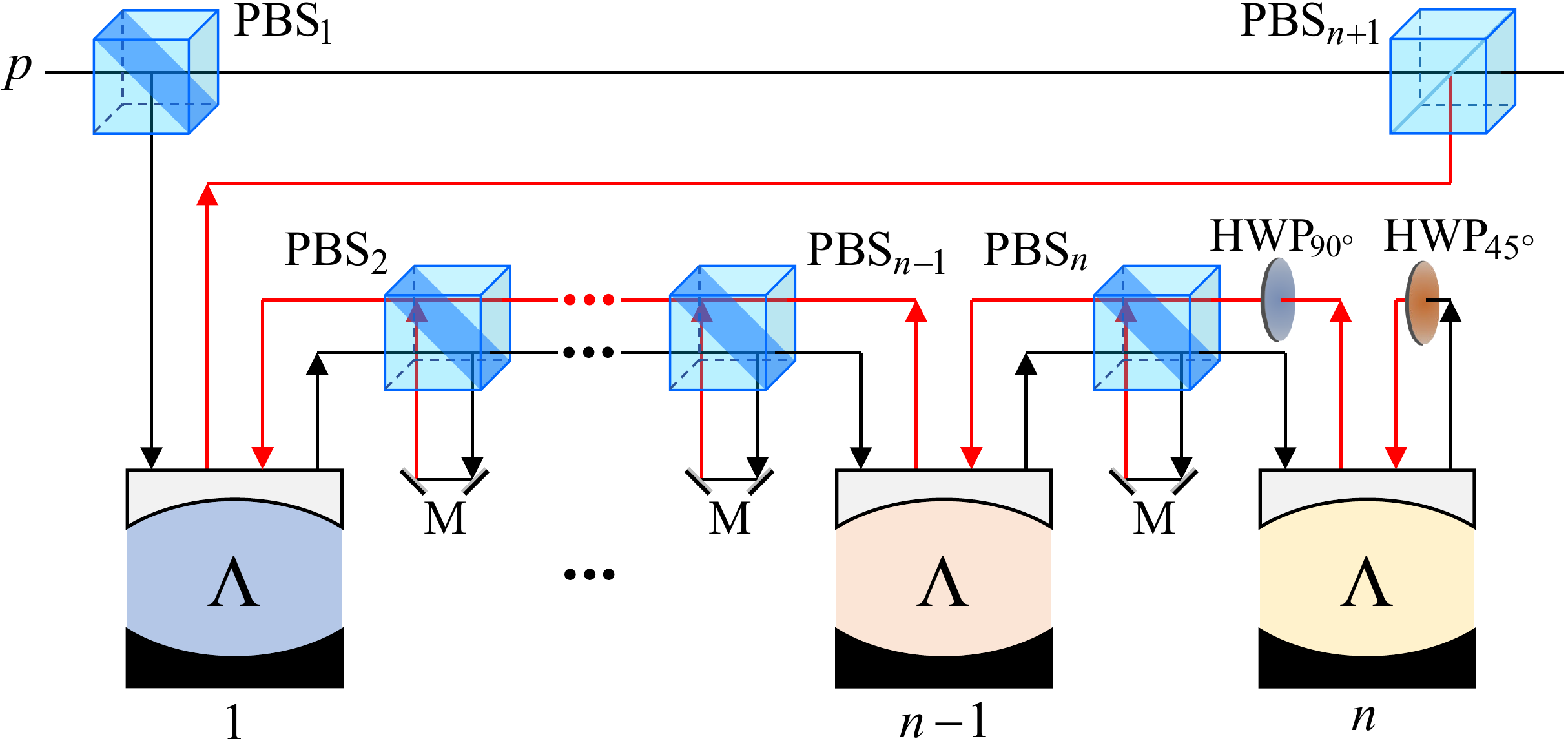}
			\label{fig5c}
	\end{minipage}}	
	\caption{Schematic diagrams of the hybrid $ n $-atom (a) CNOT, (b) Toffoli, and (c) Fredkin gates, respectively. }\label{fig5}
\end{figure}

So far, we have
established three basic logical gates for hybrid systems assisted by photons and $\Lambda$-type three-level atoms in cavities.
 Moreover, three gates only rely on the polarized state of the photon and the states of the atoms to encode information without any auxiliary particle, and that are not required to measure during the operations.
Therefore, the experimental quantum resources and the engineering complexity of the three gates in our schemes are greatly simplified.
Further, these schemes can be extended to optimize the multi-qubit hybrid CNOT, Fredkin and Toffoli gates involved with the $ n $ atoms and the photon shown in Fig. \ref{fig5} for implementing the large-scale quantum network.

Suppose that the initial states of the photon and the atoms $ 1 $, ..., $ n-2 $, $ n-1 $ and $ n $ are $ (\alpha_{1}|H\rangle+\alpha_{2}|V\rangle)_{p}$, $(\zeta_{1}|g_{h}\rangle+\zeta_{2}|g_{v}\rangle)_{1} $, ..., $(\rho_{1}|g_{h}\rangle+\rho_{2}|g_{v}\rangle)_{n-2}$, $(\upsilon_{1}|g_{h}\rangle+\upsilon_{2}|g_{v}\rangle)_{n-1}$, and $(\chi_{1}|g_{h}\rangle
	+\chi_{2}|g_{v}\rangle)_{n}$, respectively.
Specifically, we require only two PBSs and $ n $ HWP$ _{45^{\circ}} $ for implementing the hybrid multi-qubit CNOT gate shown in Fig. \ref{fig5a}, including one photon and $ n $ atoms.  If
and only if  the photon $ p $ is in the polarized state $|V\rangle$, the states of the atoms $ 1 $, $ 2 $, ..., $ n-1 $ and $ n $ simultaneously  perform   qubit-flip operations.
We note that when $ n $ is even, the hybrid $ n $-atom CNOT gate is fully implemented without extra operation. In contrast, when $ n $ is odd, it is necessary to apply a $ \sigma_{z} $ operation with the HWP$ _{0^{\circ}} $ on the polarized state $ |V\rangle $ of photon $ p $ after passing through  PBS$ _{1} $.
Therefore, the state of this hybrid $ n $-atom CNOT gate is
\begin{eqnarray}\label{eq19}
	|\Phi^{n}_{C}\rangle\!\!\!\!\!\!&=&\!\!\!\!\!\!
	\alpha_{1}|H\rangle_{p}(\zeta_{1}|g_{h}\rangle
	+\zeta_{2}|g_{v}\rangle)_{1}... \nonumber\\
&&\!\!\!\!\!\!\otimes(\upsilon_{1}|g_{h}\rangle+\upsilon_{2}|g_{v}\rangle)_{n-1}(\chi_{1}|g_{h}\rangle
	+\chi_{2}|g_{v}\rangle)_{n}\nonumber\\
	&&\!\!\!\!\!\!+(-1)^{n}\alpha_{2}|V\rangle_{p}(\zeta_{2}|g_{h}\rangle
	+\zeta_{1}|g_{v}\rangle)_{1}...\nonumber\\
	&&\!\!\!\!\!\!\otimes(\upsilon_{2}|g_{h}\rangle+\upsilon_{1}|g_{v}\rangle)_{n-1}(\chi_{2}|g_{h}\rangle
	+\chi_{1}|g_{v}\rangle)_{n}.
\end{eqnarray}

Similarly, the hybrid $ n $-atom Fredkin gate shown in Fig. \ref{fig5b} consists of $ n $ PBSs and a
HWP$ _{90^{\circ}} $.  The hybrid $ n $-atom Fredkin gate completes the function, that is, if and only if the polarized state of photon $ p $ is $ |V\rangle $, the only two atoms $ n-1 $ and $ n $ swap the qubit information with each other.
Whether $n (n > 2) $ is odd or even, the $ -\sigma_{z} $ operation by the HWP$ _{90^{\circ}} $ is applied before the photon interacts with the $ n-1 $ atom for the first time.
Then we can get the final state of the $ n $-atom Fredkin gate is
\begin{eqnarray}\label{eq20}
	|\Phi^{n}_{F}\rangle\!\!\!\!\!\!&=&\!\!\!\!\!\!
	(\alpha_{1}|H\rangle_{p}\zeta_{1}|g_{h}\rangle_{1}...\rho_{1}|g_{h}\rangle_{n-2}\nonumber\\
	&&\!\!\!\!\!\!+...+\alpha_{2}|V\rangle_{p}\zeta_{2}|g_{v}\rangle_{1}...\rho_{1}|g_{h}\rangle_{n-2})\nonumber\\
	&&\!\!\!\!\!\!\otimes(\upsilon_{1}|g_{h}\rangle+\upsilon_{2}|g_{v}\rangle)_{n-1}(\chi_{1}|g_{h}\rangle	+\upsilon_{2}|g_{v}\rangle)_{n}\nonumber\\
	&&\!\!\!\!\!\!+	\alpha_{2}|V\rangle_{p}\zeta_{2}|g_{v}\rangle_{1}...\rho_{2}|g_{v}\rangle_{n-2}\nonumber\\
	&&\!\!\!\!\!\!\otimes(\chi_{1}|g_{h}\rangle+\chi_{2}|g_{v}\rangle)_{n-1}(\upsilon_{1}|g_{h}\rangle	+\upsilon_{2}|g_{v}\rangle)_{n}.
\end{eqnarray}

The hybrid $ n $-atom Toffoli gate shown in Fig. \ref{fig5c} consists of $ n+1 $ PBSs, one HWP$ _{90^{\circ}} $ and one HWP$ _{45^{\circ}} $.
It completes the function that if and only if the polarized state of photon $ p $ is $ |V\rangle $ and the state of the atom $ n-1 $ is $|g_{v}\rangle_{n-1}$, only the atom $ n $ performs qubit-flip operation.
Before and after the photon $ p $ interacts with the last atom  $ n $ for the first and second times, the photon in sequence passes through the HWP$ _{45^{\circ}} $ and HWP$ _{90^{\circ}} $ to perform the $ \sigma_{x} $ and $ -\sigma_{z} $ operations, respectively.
The final state of the $ n $-atom system is changed into
\begin{eqnarray}\label{eq21}
	|\Phi^{n}_{T}\rangle\!\!\!\!\!\!&=&\!\!\!\!\!\!
	(\alpha_{1}|H\rangle_{p}\zeta_{1}|g_{h}\rangle_{1}...\rho_{1}|g_{h}\rangle_{n-2}\upsilon_{1}|g_{h}\rangle_{n-1} \nonumber\\
&&\!\!\!\!\!\!+...+\alpha_{2}|V\rangle_{p}\zeta_{2}|g_{v}\rangle_{1}...\rho_{1}|g_{v}\rangle_{n-2}\upsilon_{1}|g_{h}\rangle_{n-1})\nonumber\\
&&\!\!\!\!\!\!\otimes(\chi_{1}|g_{h}\rangle	+\chi_{2}|g_{v}\rangle)_{n}\nonumber\\
&&\!\!\!\!\!\!	+(\alpha_{2}|V\rangle_{p}\zeta_{2}|g_{v}\rangle_{1}...\rho_{1}|g_{v}\rangle_{n-2}\upsilon_{1}|g_{v}\rangle_{n-1})\nonumber\\
&&\!\!\!\!\!\!\otimes(\chi_{1}|g_{v}\rangle	+\chi_{2}|g_{h}\rangle)_{n}.
\end{eqnarray}

\begin{figure}[htp]
	\centering
	\subfigure[]{
		\begin{minipage}{0.48\linewidth}
			\centering
			\includegraphics[width=1\linewidth]{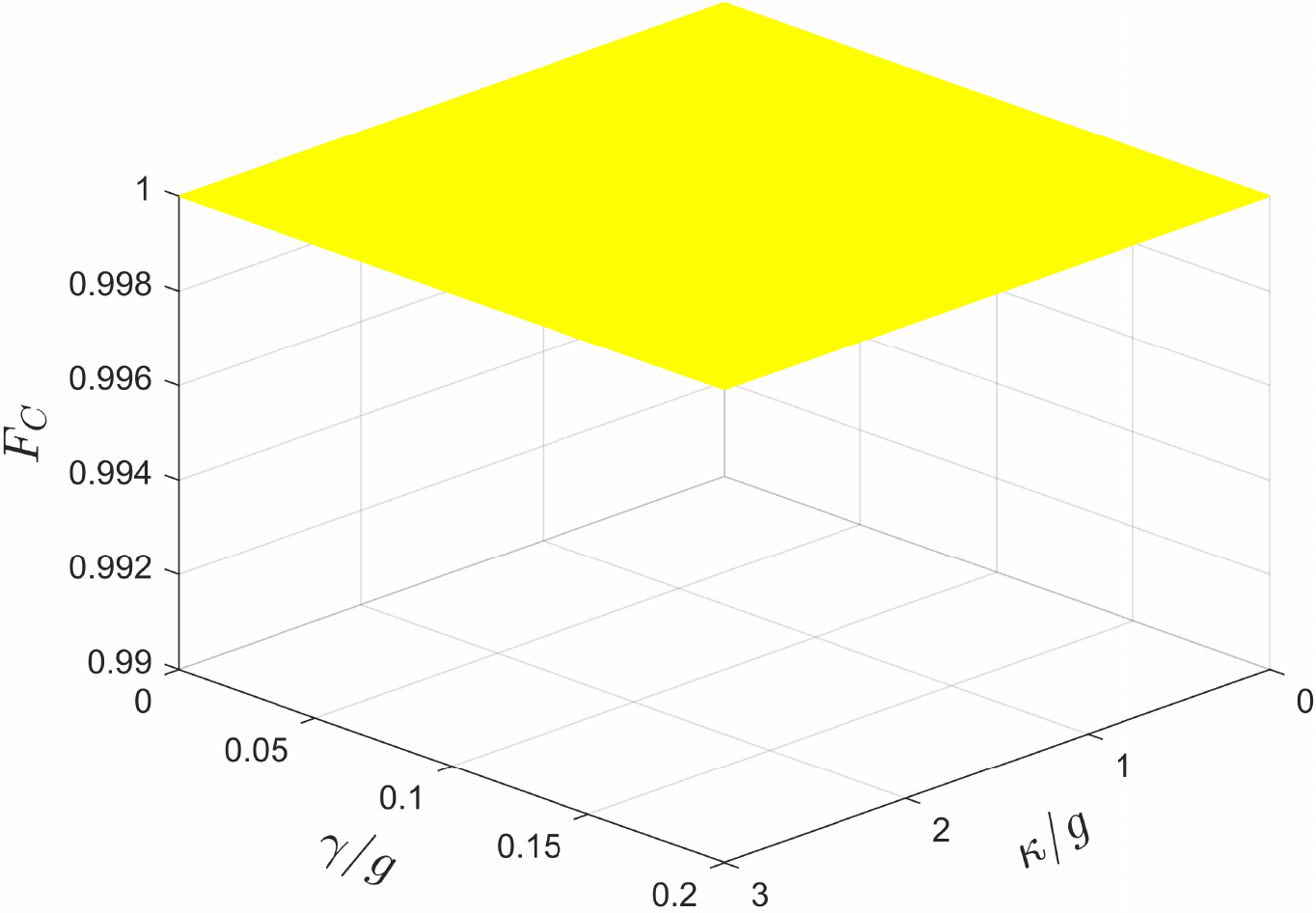}
			\label{fig6a}
	\end{minipage}}	
	\subfigure[]{
		\begin{minipage}{0.48\linewidth}
			\centering
			\includegraphics[width=1\linewidth]{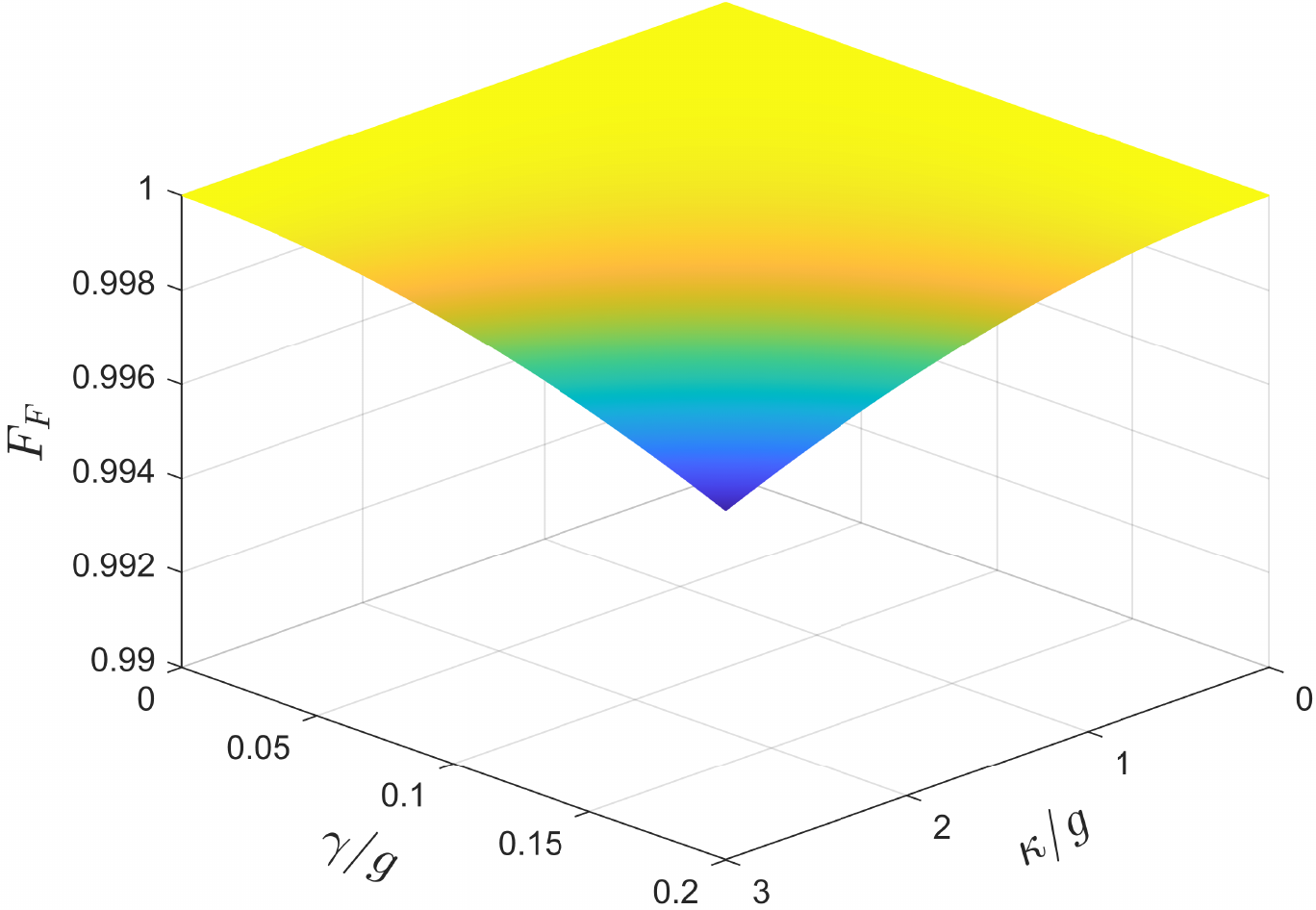}
			\label{fig6b}
	\end{minipage}}
	\subfigure[]{
		\begin{minipage}{0.48\linewidth}
			\centering
			\includegraphics[width=1\linewidth]{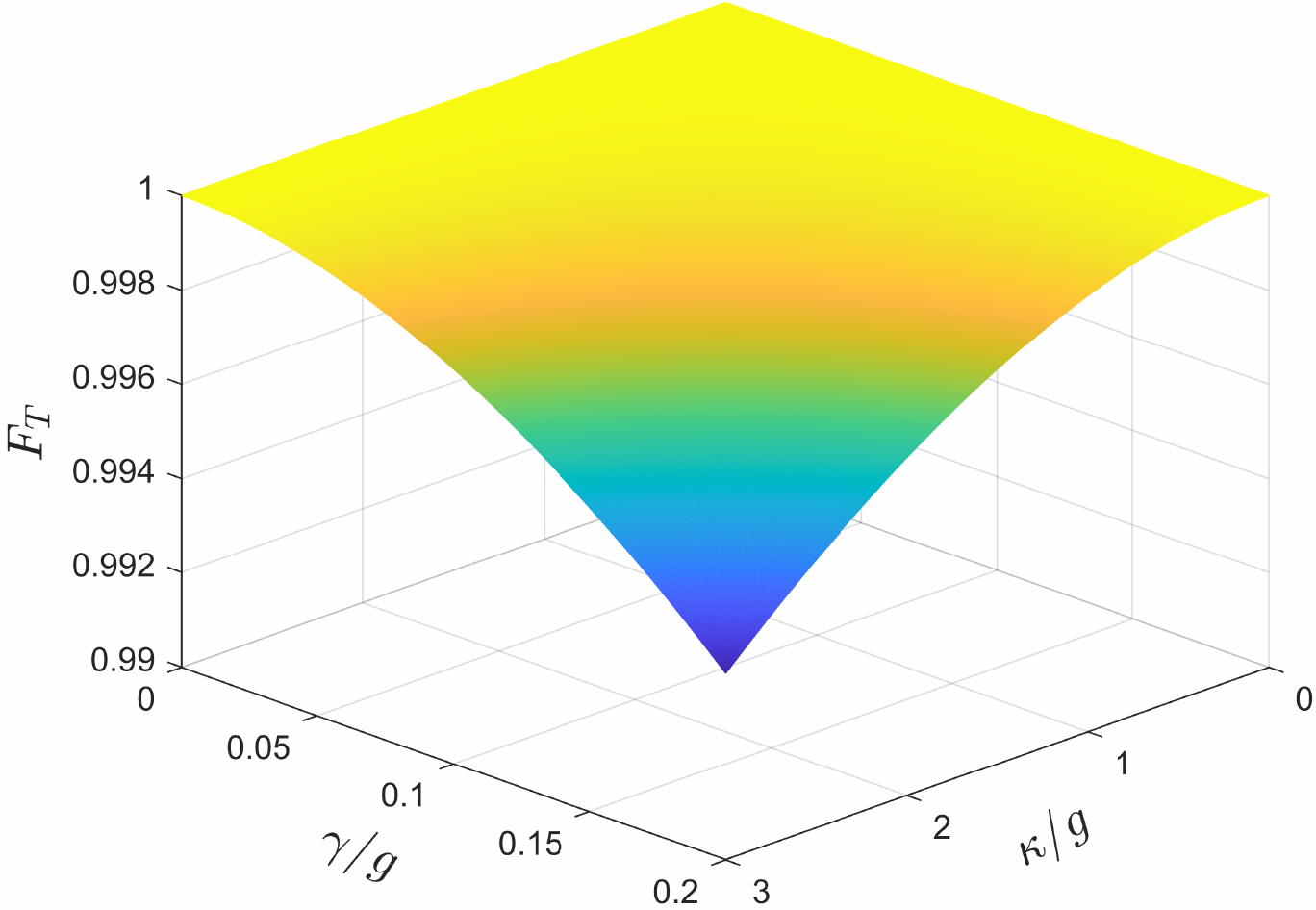}
			\label{fig6c}
	\end{minipage}}	
	\caption{The fidelities  of (a) the CNOT gate, (b) Fredkin gate,  and (c) Toffoli gate as the function of $ \gamma/g$ and $ \kappa/g $ at $ \eta_{k}=\eta_{\bar{k}}=g $ and $ \omega=0 $. }\label{fig6}
\end{figure}

We have constructed three multi-qubit gates without any auxiliaries and all the single-qubit operations are applied on the photon only, which effectively reduces running time and complexity. Besides, the decoherence effect of the atom-photon hybrid system should be considered, that is, when the photon interacts with the same atom for the second time, they should maintain the coherence time.
Next we present a detail discussion about the experimental
implementations of the CNOT, Fredkin, and Toffoli gates on hybrid systems.
The level configuration of the atom in Fig. \ref{fig1} can be found in
$^{87}Rb$ \cite{Cavity2004}, that is, the levels  $S_{1/2}$ and $P_{3/2}$ of $^{87}Rb$ atom act as the ground state and the excited state, respectively. And when $^{87}Rb$ is trapped at the center of an optical cavity, the long trapping time (tens of seconds)
of the atom in the cavity can be obtained in Ref. \cite{Ground-State2013}, so the atoms can be considered as
a good stationary qubits. 
And the fidelities and efficiencies of the three logical gates can be calculated according to the rules of actual interactions between photon and the atom-cavity system as follows
\begin{eqnarray}\label{eq22}
	|H\rangle|g_{h}\rangle&\rightarrow&r_{h1}|H\rangle|g_{h}\rangle+r_{h2}|V\rangle|g_{v}\rangle,\nonumber\\
	|H\rangle|g_{v}\rangle&\rightarrow&r_{0}|H\rangle|g_{v}\rangle,\nonumber\\
	|V\rangle|g_{h}\rangle&\rightarrow&r_{0}|V\rangle|g_{h}\rangle,\nonumber\\
	|V\rangle|g_{v}\rangle&\rightarrow&r_{h2}|H\rangle|g_{h}\rangle+r_{h1}|V\rangle|g_{v}\rangle.
\end{eqnarray}
For simplifying the calculations of the fidelity and efficiency, the injected photon is assumed resonantly coupled to the atom-cavity system (i.e., $ \omega=0 $), and the photon, atoms are in maximum entangled initial state at the same time.
Besides, the coupling strength is taken as $ \eta_{k}=\eta_{\bar{k}}=g $, as a result, the two parameters $ \kappa/g$ and $ \gamma/g $ become the key factors affecting the fidelity and efficiency.
As shown in Figs. \ref{fig6}(a)-(c), one can see that
the fidelity of CNOT gate is near-unit without the affect of both the parameters, and the fidelity of  Fredkin and Toffoli gates are further improved with the decrease of both the parameters.
For example, the fidelities of Fredkin and Toffoli gates are improved from $F_{F}=99.88\%$ to $F_{F}=99.97\%$, and
$F_{T}=99.72\%$ to $F_{T}=99.93\%$, respectively, as the parameter decreases from $\kappa/g=2$ to $\kappa/g=1$ at the parameter $\gamma/g=0.2$.

\begin{figure}[htp]
	\centering
	\subfigure[]{
		\begin{minipage}{0.48\linewidth}
			\centering
			\includegraphics[width=1\linewidth]{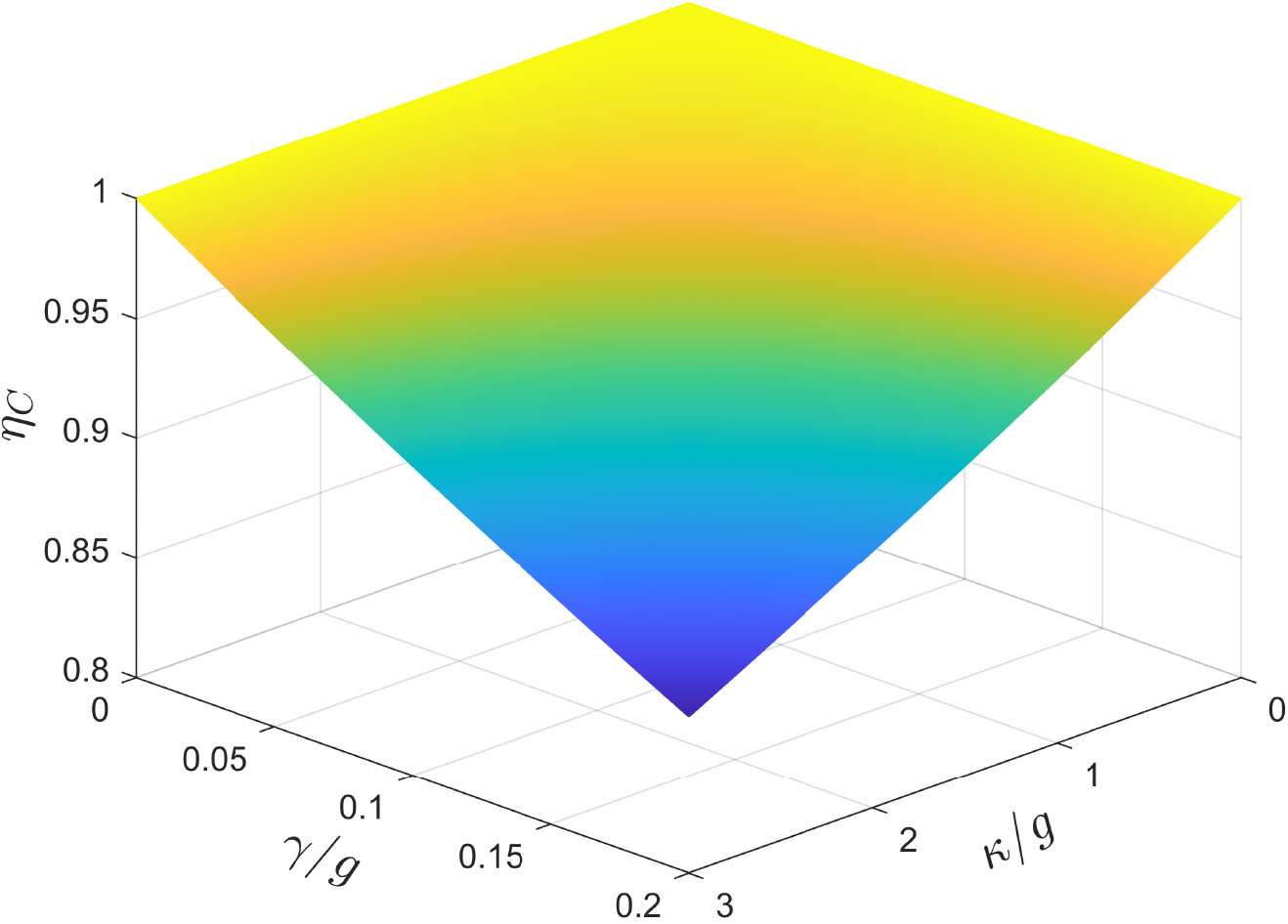}
			\label{fig7a}
	\end{minipage}}	
	\subfigure[]{
		\begin{minipage}{0.48\linewidth}
			\centering
			\includegraphics[width=1\linewidth]{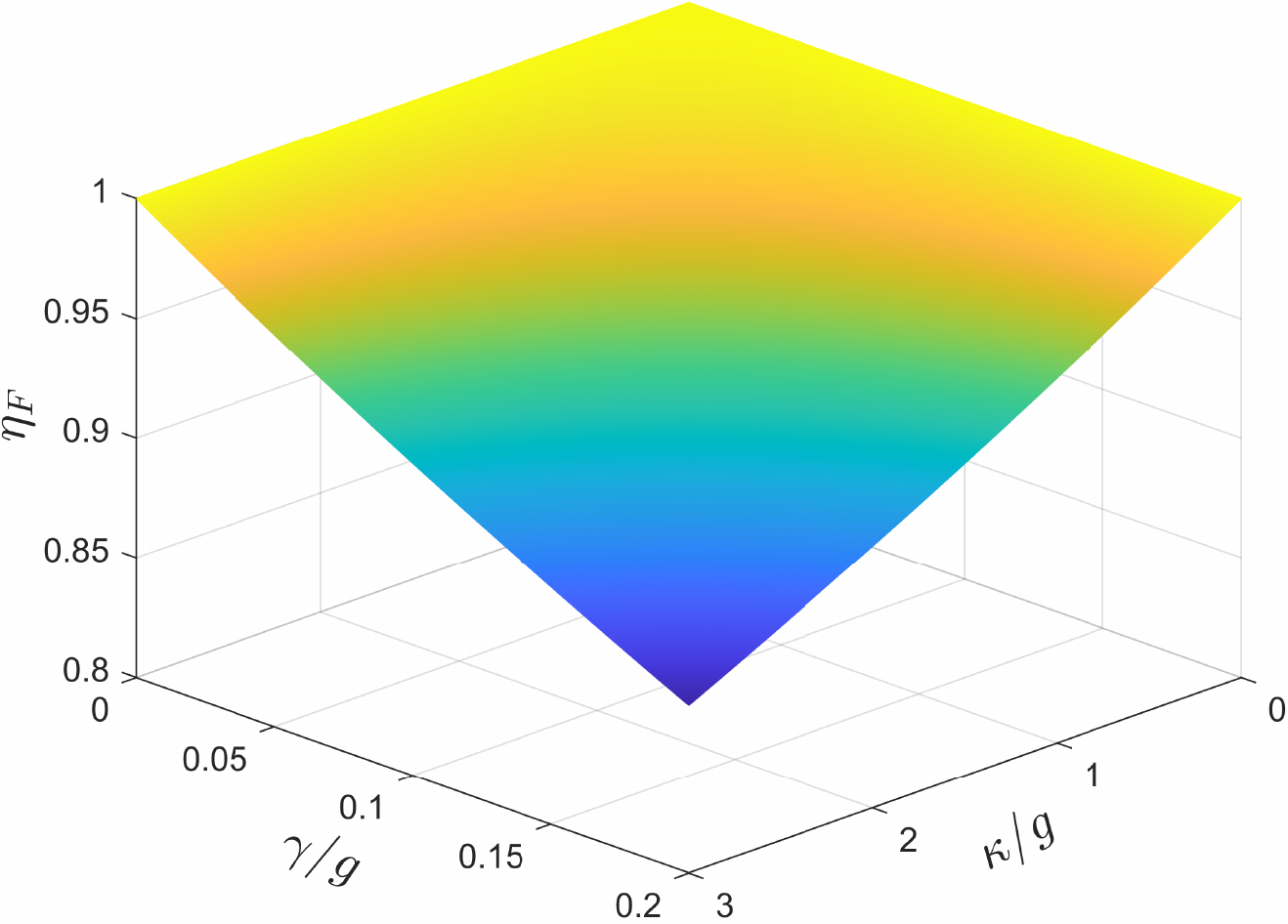}
			\label{fig7b}
	\end{minipage}}
	\subfigure[]{
		\begin{minipage}{0.48\linewidth}
			\centering
			\includegraphics[width=1\linewidth]{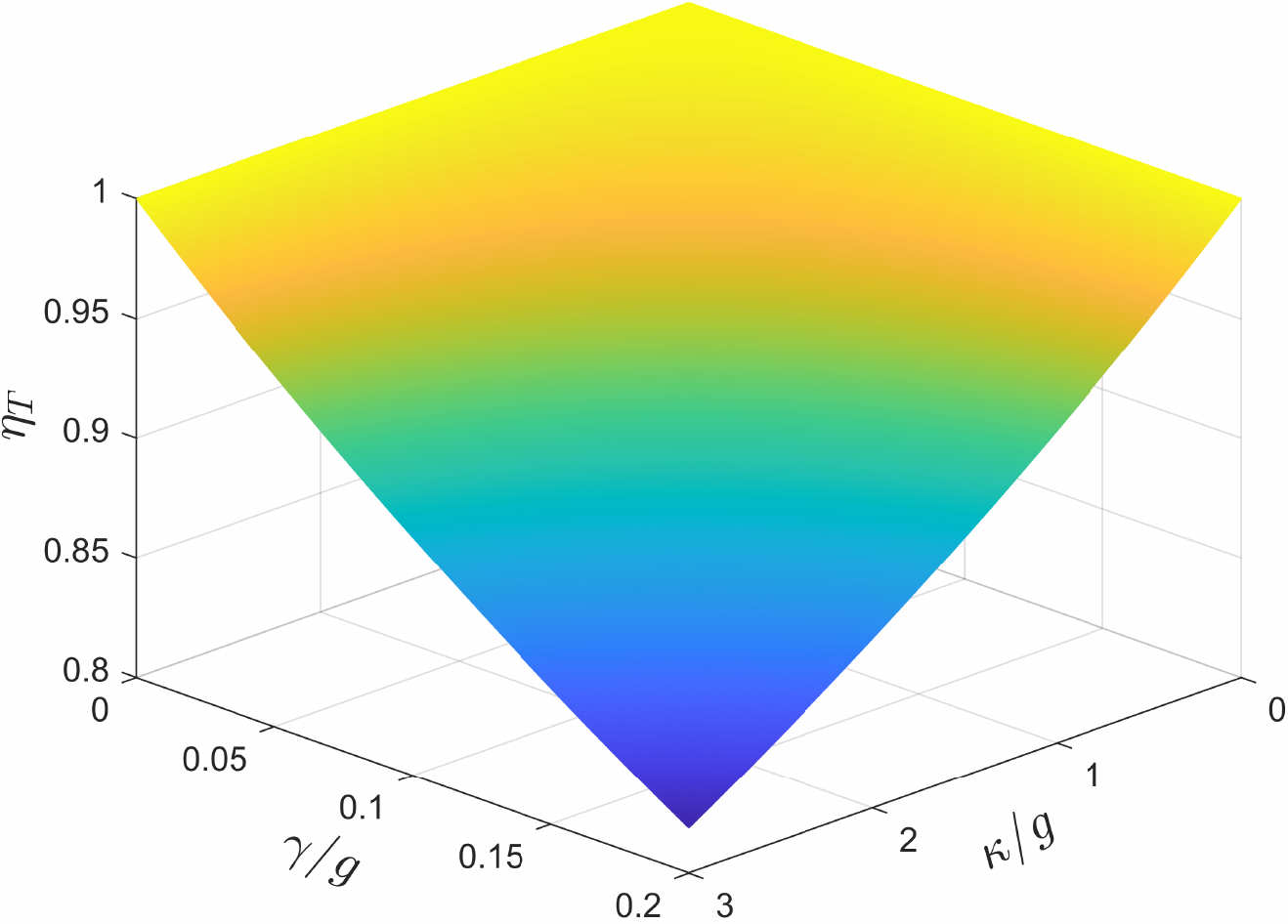}
			\label{fig7c}
	\end{minipage}}	
	\caption{The efficiencies of (a) the CNOT gate, (b) Fredkin gate,  and (c) Toffoli gate as the function of $ \gamma/g$ and $ \kappa/g $ at $ \eta_{k}=\eta_{\bar{k}}=g $ and $ \omega=0 $. }\label{fig7}
\end{figure}

Moreover, the efficiency, i.e., the other feature, is the ratio of the output state to the input state of the photon.
Similarly,
it can be seen from Figs. \ref{fig7}(a)-(c) that the efficiencies of the three gates are further improved with the decrease of both the parameters.
For example, the efficiencies of CNOT, Fredkin and Toffoli gates are respectively improved from $\eta_{C}=95.18\%$ to $\eta_{C}=97.55\%$, $\eta_{F}=95.24\%$ to $\eta_{F}=97.56\%$ and
$\eta_{T}=93.04\%$ to $\eta_{T}=96.39\%$ as the parameter decreases from $\gamma/g=0.2$ to $\gamma/g=0.1$ at the parameter $\kappa/g=1$.

Combined with the existing experimental data, the simulations indicate that these schemes are insensitive
to cavity decay and atomic spontaneous emission, so they could work under the condition of a larger cavity decay rate,
i.e., a cavity with a relatively lower-$Q$ factor.
Finally, we must point out that it is still a high challenge
to construct an array of atom-cavity systems to implement
multi-qubit quantum gates with the present technology.
However, we believe that the rapid development in the
field of cavity QED will offer the useful technology
for large-scale quantum computing in the
future.

In summary, we have proposed three refined protocols for realizing CNOT, Fredkin, and Toffoli gates on atom-photon hybrid systems by the reflection geometry of a single photon interacting with a three-level $\Lambda$-typle atom-cavity system.
These quantum gates can be flexibly extended  to  multi-qubit CNOT, Fredkin and Toffoli gates with $ O(n) $ optical elements, and none of the above gates require auxiliary particles.
In addition, all the single-qubit operations are applied on the photon only.
Our schemes have high fidelities in weak coupling regime, possessing relatively long coherence time, which are much different from strong-coupling case, and having potential applications on scalable quantum computing.
Therefore, it is more feasible to realize not only fast quantum operations, but also multi-time operations between photon and the cavity-atom system.

\acknowledgments
This work was supported in part by the Natural Science Foundation of China under Contract 61901420;	
in part by the Shanxi Province Science Foundation for Youths under Contract 201901D211235;
in part by the Scientific and Technological Innovation Programs
of Higher Education Institutions in Shanxi under Contract 2019L0507;
in part by the Shanxi "1331 Project" Key Subjects Construction.

\end{document}